\documentclass[12pt,preprint]{aastex}
\usepackage{emulateapj5}
\usepackage{apjfonts}
\usepackage{epsfig}
\usepackage{natbib}
\newcommand{\ciii}{\ion{C}{3}]}
\newcommand{\civ}{\ion{C}{4}}

\newcommand{\flamb}{erg s$^{-1}$ cm$^{-2}$ \AA$^{-1}$}

\newcommand{\feii}{\ion{Fe}{2}}
\newcommand{\feiii}{\ion{Fe}{3}}

\def\gtrsim{\mathrel{\hbox{\rlap{\hbox{\lower4pt\hbox{$\sim$}}}\hbox{\raise2pt\hbox{$>$}}}}}

\newcommand{\fwha}{\ensuremath{\mathrm{FWHM}_\mathrm{H{\alpha}}}}
\newcommand{\fwhb}{\ensuremath{\mathrm{FWHM}_\mathrm{H{\beta}}}}

\newcommand{\halpha}{H\ensuremath{\alpha}}

\newcommand{\hbeta}{H\ensuremath{\beta}}

\newcommand{\hn}{\halpha+[\ion{N}{2}]}
\newcommand{\hst}{\emph{HST}}

\newcommand{\kms}{km~s\ensuremath{^{-1}}}

\newcommand{\lf}{\ensuremath{L_{\rm{5100 \AA}}}}

\newcommand{\lum}{erg s$^{-1}$}

\newcommand{\lledd}{\ensuremath{L_{\mathrm{bol}}/L{\mathrm{_{Edd}}}}}

\newcommand{\mbulge}{\ensuremath{M_\mathrm{BH}-L_{\mathrm{bulge}}}}

\newcommand{\mbh}{\ensuremath{M_\mathrm{BH}}}

\newcommand{\mgii}{\ion{Mg}{2}}

\newcommand{\msigma}{\ensuremath{M_{\mathrm{BH}}-\sigmastar}}
\newcommand{\msun}{\ensuremath{M_{\odot}}}

\newcommand{\nii}{[\ion{N}{2}]}

\newcommand{\oiii}{[\ion{O}{3}]}

\newcommand{\sers}{S{\'e}rsic}

\newcommand{\sigmastar}{\ensuremath{\sigma_{\ast}}}

\newcommand{\zw}{1~Zw~{\small I}}

\def\lax{{$\mathrel{\hbox{\rlap{\hbox{\lower4pt\hbox{$\sim$}}}\hbox{$<$}}}$}}
\def\gax{{$\mathrel{\hbox{\rlap{\hbox{\lower4pt\hbox{$\sim$}}}\hbox{$>$}}}$}}

\slugcomment{Accepted for publication in {\it The Astrophysical Journal}.}
\shortauthors{GREENE ET AL.}

\begin{document}

\title{Redshift Evolution in Black Hole-Bulge Relations: Testing 
C{\small IV}-based Black Hole Masses}

\author{Jenny E. Greene}
\affil{Department of Astrophysical Sciences, Princeton University, 
Princeton, NJ 08544; Russell, Princeton-Carnegie Fellow}

\author{Chien Y. Peng}
\affil{NRC Herzberg Institute of Astrophysics, 5071 West Saanich Road, 
Victoria, BC V9E2E7, Canada}

\author{Randi R. Ludwig}
\affil{University of Texas at Austin, Department of Astronomy, 
1 University Station, C1400 Austin, TX 78712}

\begin{abstract}

We re-examine claims for redshift evolution in black hole-bulge
scaling relations based on lensed quasars.  In particular, we refine
the black hole mass estimates using measurements of Balmer lines
from near-infrared spectroscopy obtained with Triplespec at Apache
Point Observatory.  In support of previous work, we find a large
scatter between Balmer and UV line widths, both \mgii$~\lambda
\lambda 2796, 2803$ and \civ$~\lambda \lambda 1548, 1550$.  There is
tentative evidence that \ciii$~\lambda 1909$, despite being a blend
of multiple transitions, may correlate well with \mgii, although a
larger sample is needed for a real calibration.  Most importantly,
we find no systematic changes in the estimated BH masses for the
lensed sample based on Balmer lines, providing additional support to
the interpretation that black holes were overly massive compared to
their host galaxies at high redshift.

\end{abstract}


\section{Evolution in Black Hole-Bulge Relations}

Locally, we observe tight correlations between the properties of
bulge-dominated galaxies and the masses of their central supermassive
black holes \citep[BHs; e.g.,][]
{kormendyrichstone1995,tremaineetal2002, gultekinetal2009}.  The
mechanisms that establish and maintain these relations are uncertain,
despite innumerable suggestions in the literature
\citep[e.g.,][]{silkrees1998,murrayetal2005,
  miraldakollmeier2005,hopkinsetal2006,peng2007}.  In principle,
evaluating the demographics of nuclear BHs as a function of redshift
should observationally constrain the processes that lead to the tight
scaling relations observed today.  Unfortunately, it is currently
prohibitive to obtain dynamical BH masses for systems beyond tens of
Mpc.  Thus, estimates of BH mass at large distance are necessarily
based on very indirect methods linked to accretion processes in active
BHs \citep[e.g.,][]{vestergaard2002}.

Several studies have used active galaxies to probe evolution in
BH-bulge scaling relations, probing redshifts from $0.4 < z < 0.6$
\citep{wooetal2006,treuetal2007,wooetal2008} and 1$\lesssim z
\lesssim$ 4 \citep{pengetal2006a,pengetal2006b,salvianderetal2007,
  shieldsetal2006,ho2007,jahnkeetal2009,mcleodbechtold2009} all the
way to $z=6.4$ \citep{walteretal2004}.  Generally speaking, a wide
variety of observations suggest that BH-bulge relations do evolve with
redshift, in the sense that the ratio of BH to bulge mass was higher
at early times \citep[but see
also][]{shieldsetal2003,alexanderetal2005}, although we have
observational constraints only for the most massive systems
(\mbh$>10^8$~\msun) at high redshift.  This counterintuitive result
has stimulated vigorous discussion both about the ramifications for
the coevolution of BHs and bulges
\citep[e.g.,][]{robertsonetal2006,croton2006} and about whether there
are built-in biases in the measurement techniques
\citep{laueretal2007}.

Unfortunately, even apart from potential population biases, our
interpretation of the observations are prone to significant
uncertainty.  On the one hand, it is only possible to obtain BH mass
estimates at cosmological distances using active galaxies (typically
luminous quasars at high redshift).  Immediately it becomes very
challenging to characterize the host galaxy properties, when the
quasar outshines the underlying galaxy starlight by factors of $\sim
10-30$ \citep[e.g.,][]{pengetal2006a,kimetal2008a}.  Apart from direct
imaging, some groups have used gas measurements (predominantly CO) to
obtain dynamical masses \citep{shieldsetal2006,hoetal2008b,walteretal2004,riechersetal2008,riechersetal2009}, which may or may
not provide a reliable tracer of the galaxy mass \citep{ho2007}.  The
width of narrow emission lines, particularly \oiii~$\lambda 5007$,
have also been substituted for the galaxy velocity dispersion
\citep[e.g.,][]{shieldsetal2003,boroson2003,salvianderetal2007,gaskell2009}.
While there is a strong correlation between stellar and gaseous
velocity dispersion in low-luminosity sources
\citep[e.g.,][]{heckmanetal1981,nelsonwhittle1996,greeneho2005o3,ho2009o3},
there is good reason to suspect that it does not hold at high
luminosity \citep[e.g.,][]{greeneetal2009}.  \citet[][P06
hereafter]{pengetal2006b} mitigated the host galaxy contrast
problem by focusing on lensed quasars.  The quasars are lensed
differently from the underlying (resolved) host galaxies, reducing the
contrast problem discussed above.  For that reason, we focus on the
lensed quasar sample in this paper.

Daunting as measuring high-redshift galaxy mass and stellar velocity
dispersion may be, particularly in the presence of a luminous quasar,
the BH mass measurements are equally problematic.  The techniques are
indirect and model-dependent.  Briefly, active galaxies contain dense
gas orbiting at distances of light days to months from the central BH
that gives rise to broad emission lines with widths of thousands of
\kms.  By combining a size scale with the line width of the emitting
region, the broad-line region (BLR) gas can be used as a dynamical
tracer of the BH mass \citep[e.g.,][]{dibai1980}.  Direct size
estimates are obtained by measuring the time lag between variability
in the continuum and line emission \citep[reverberation
mapping;][]{blandfordmckee1982,petersonetal2004}.  
\hskip 0.in
\psfig{file=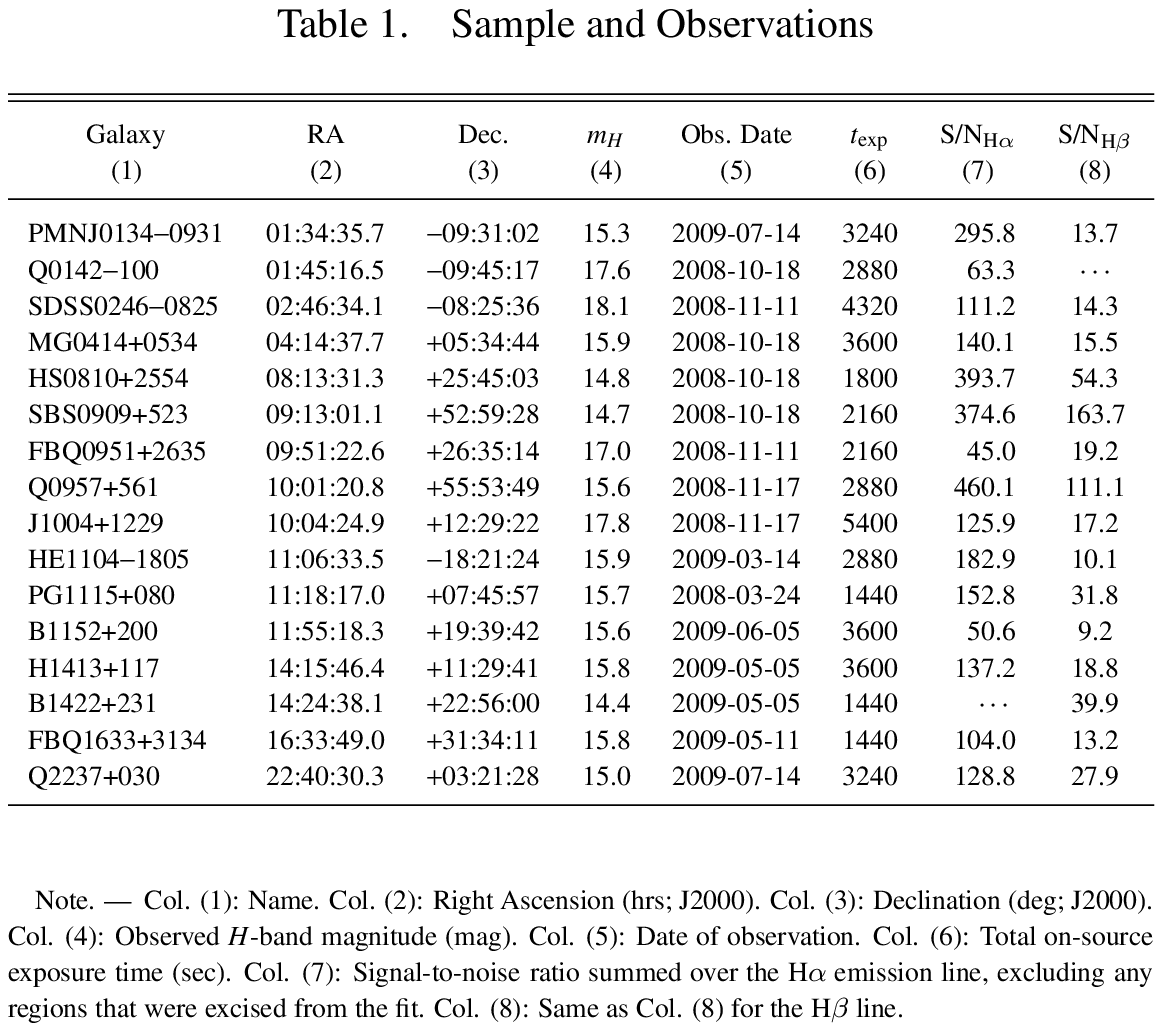,width=0.45\textwidth,keepaspectratio=true,angle=0}
\vskip 4mm
\noindent
A tight, empirically determined correlation between BLR radii and the
luminosity of the active galaxy \citep[the radius-luminosity relation;
][] {kaspietal2005,bentzetal2006,bentzetal2009a} can be used to obtain
approximate BLR radii for the quasar population in general
\citep[e.g.,][]{vestergaard2002}.  The current data support a slope of
$R_{\rm BLR} \propto L^{0.5}$, as is expected if neither the density
structure of the BLR nor the shape of the ionizing continuum depends
on luminosity \citep[e.g.,][]{bentzetal2006}.  Unfortunately, the
radius-luminosity relation has not been calibrated extensively for
luminosities greater than \lf$\approx 10^{46}$~erg~s$^{-1}$
\citep[see][]{kaspietal2007}.

Because the kinematic structure and inclination of the BLR are
unknown, the derived virial ``masses'' (\mbh$\propto \upsilon^2 R_{\rm
  BLR}/G$) have no physically motivated normalization.  Recent
reverberation mapping experiments reveal signatures of inflow,
rotation, and outflow in individual objects, but such two-dimensional
maps remain scarce
\citep{welshhorne1991,bentzetal2009b,denneyetal2009rm}.  Our current
practice is to compare the virial masses with independent estimates of
BH mass, typically using the \msigma\ relation, to derive an average
scale factor
\citep[e.g.,][]{gebhardtetal2000b,ferrareseetal2001,onkenetal2004,
  nelsonetal2004,greeneho2006msig,shenetal2008a}.  The fact that any
correlation is seen between virial masses and the mass inferred from
the \msigma\ relation is encouraging, but leaves much room for large
systematic uncertainties \citep[e.g.,][]{krolik2001,collinetal2006}.

At higher redshift, only rest-frame UV spectra are readily available
for large samples of quasars.  The scaling relations for the UV lines
(e.g.,~\mgii~$\lambda 2800$~\AA; \civ~$\lambda 1550$~\AA) include
additional layers of uncertainty.  Virtually no reverberation mapping
has been performed with \mgii, and the scaling relations for this line
are simply scaled to match \hbeta\
\citep[e.g.,][]{mcluredunlop2004,onkenkollmeier2008}.  In the case of
\civ, reverberation mapping has been done
\citep[e.g.,][]{petersonetal2005}, but there are strong reasons to
suspect that the \civ\ line width is not dominated by virial motions
\citep[e.g.,][]{gaskell1982,baldwinetal1996,richardsetal2002a,leighlymoore2004,
  baskinlaor2005,sulenticetal2007,shenetal2008b} although debate
continues on this point
\citep[e.g.,][]{vestergaardpeterson2006,kellybechtold2007,
  gavignaudetal2008}.  For these reasons, virial masses based on
Balmer lines \citep[preferably \halpha;][]{greeneho2005cal} have the
most credibility, since these have been directly compared with
alternate estimates of \mbh.  We focus specifically on obtaining
Balmer-based virial masses for the high-redshift lensed quasar sample
from P06.  Our primary goal is to determine whether the masses
presented in P06 are {\it systematically} biased by the use of UV line
transitions.  We follow P06 and assume a standard cosmology with $H_0
= 100~h = 70$~\kms~Mpc$^{-1}$, $\Omega_{\rm m} = 0.30$, and
$\Omega_{\Lambda} = 0.70$.

\section{Observations and Data Reduction}

The data presented here were obtained over the course of a year using
the newly commissioned near-infrared spectrograph Triplespec
\citep{wilsonetal2004} at Apache Point Observatory (Table 1).  All
objects were observed with a $1.1 \times 43\arcsec$ slit and Fowler
sampling of 8 \citep[the number of non-destructive readouts at the
beginning and end of each exposure designed to minimize
readnoise;][]{fowlergatley1990}.  Triplespec covers a nominal
wavelength range of 0.95-2.46 $\micron$ with $R = 3500$. Observing
conditions ranged from clear to partly cloudy, with typical seeing of
$\theta \approx 1\farcs 5$.  In most cases the slit was positioned at
the parallactic angle in the middle of the observation, although in a
couple of cases we positioned the slit to place two quasar images in
the slit at once.  The object was dithered along the slit every 180
sec to improve sky subtraction.  For each quasar we observed a nearby
A0V star ($10 < H < 6$ mag) to serve as flux and telluric standard.

\begin{figure*}
\vbox{ 
\vskip -3mm
\hskip -0.in
\psfig{file=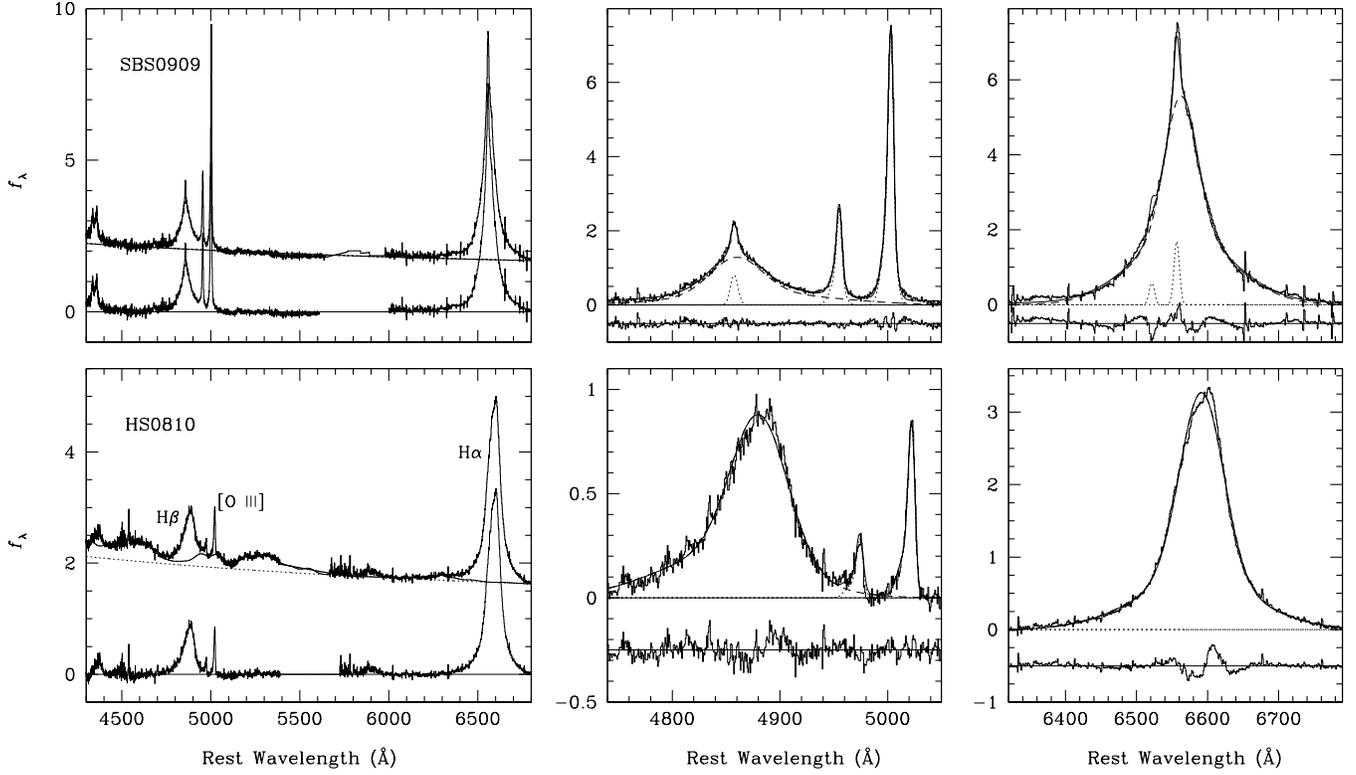,width=0.55\textwidth,keepaspectratio=true,angle=90}
}
\vskip -0mm
\figcaption[]{
Example fits to the continuum ({\it left}), \hbeta\ ({\it middle}), 
and \halpha\ ({\it right}) lines from the Triplespec data.  We show the 
data ({\it solid histogram}), the total model ({\it thin solid}), 
the broad- ({\it dashed}) and narrow-line ({\it dotted}) model 
components, and residuals below ({\it thin solid histogram}).  Data are 
plotted with an arbitrary scale in $f_{\lambda}$.  The rest of the sample 
are plotted in the Appendix.
\label{fits}}
\end{figure*}

The data were reduced using custom software that is a modified version
of Spextool and is described in detail in \citet{vaccaetal2003} and
\citet{cushingetal2004}.  Using dome-flat and arc-line exposures, the
code creates and applies a flat-field and wavelength solution.  Bias
and dark subtraction is accomplished through pair-wise differencing of
images taken at two slit positions, which also removes air glow
emission from the atmosphere, at least to zeroth order.  Nonlinearity
corrections are applied and then each pair of spectra are traced and
optimally extracted \citep{horne1986}, including background
subtraction.  Wavelength calibration is applied, and all the spectra
of a given source are median-combined.  Flux calibration is
accomplished using an A0V star observed at similar time and airmass.
This same star is used to create a model of the telluric absorption by
assuming that the A star has an intrinsic spectrum identical to that
of Vega (see Vacca et al. for details).  Prior to correction, small
wavelength shifts between the A star and the program object are
derived on an order by order basis using a cross-correlation
technique.  Finally the orders are merged with small scale factors
applied to properly match the edges of each order, and cosmetic data
clipping is done.  The software also generates an error array.
Heliocentric corrections are calculated using the IRAF task {\it
  bcvcorr}.

\begin{figure*}
\vbox{ 
\vskip -3mm
\hskip 0.in
\psfig{file=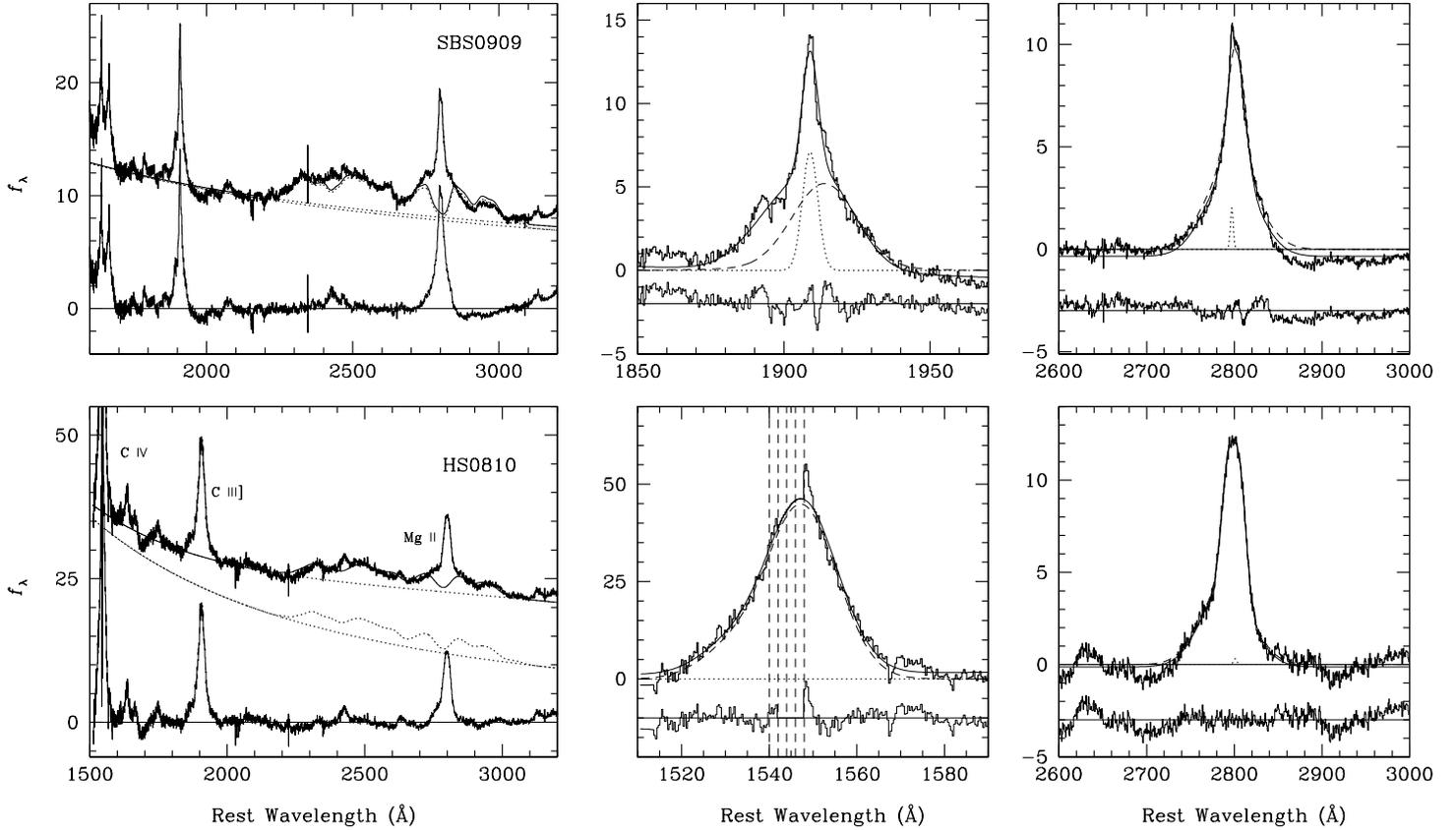,width=0.6\textwidth,keepaspectratio=true,angle=90}
}
\vskip -0mm
\figcaption[]{
Example fits to the continuum ({\it left}), C {\tiny IV} or 
C {\tiny III}] ({\it
  middle}), and Mg {\tiny II} ({\it right}) lines from the SDSS data, shown
in units of $10^{-15}$~\flamb.
We show the data ({\it solid histogram}), the total model ({\it thin
  solid}), the broad- ({\it dashed}) and narrow-line ({\it dotted})
model components, and residuals below ({\it thin solid histogram}).
The dashed vertical lines denote a masked region of the spectrum
that is heavily absorbed in the original spectrum.  Note that in the
C {\tiny III}] profile ({\it top-middle panel}) we show only the broad
component of the C {\tiny III}] line itself.
\label{fits}}
\end{figure*}

The resulting S/N for each object across the \halpha\ and \hbeta\
lines are shown in Table 1.  The relative flux calibration is
reasonable and the resulting spectra have smooth, power-law continua.
However, given the variable clouds that plagued many of our
observations, we suspect that the overall flux calibration scale is
not reliable.  For instance, the flux scales for two observations of
B1152+200 differ by a factor of three.  Our results are not impacted
by these problems, however, since we measure intrinsic luminosities
from broad-band photometry combined with a lens model (P06 and see below).

\subsection{Rest-frame Ultraviolet Spectra from SDSS}

In addition to the rest-frame optical spectra obtained with
Triplespec, we also utilize observed optical (rest-frame ultraviolet;
UV) spectra from the the Sloan Digital Sky Survey
\citep[SDSS;][]{yorketal2000,abazajianetal2009}.  These targets were
selected from the SDSS photometry as quasars
\citep{richardsetal2002b}.  Galaxies with SDSS observations have
tabulated UV slopes in Table 2.

\section{Continuum and Line Measurements}

\begin{figure*}
\vbox{ 
\hskip 0.6in
\psfig{file=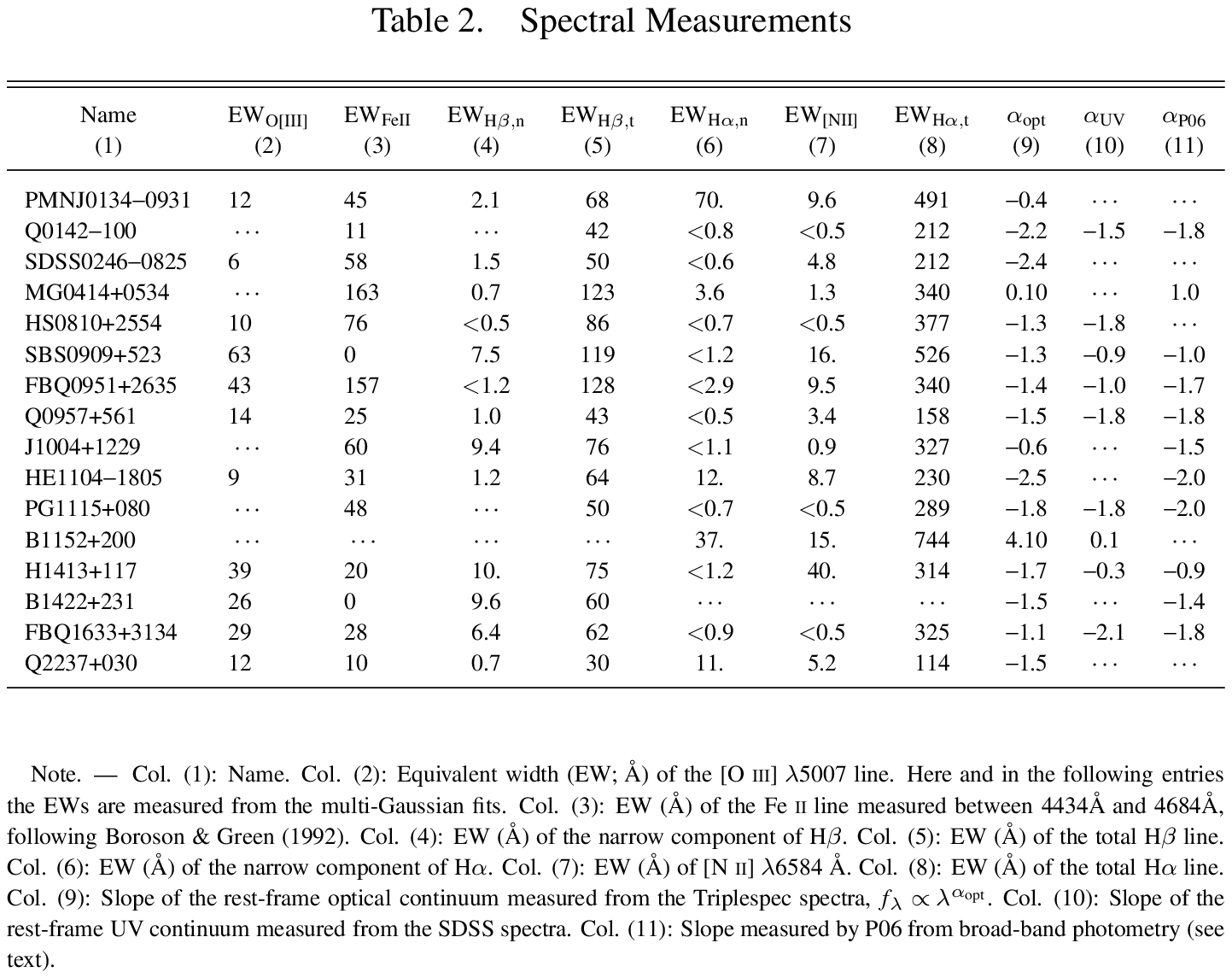,width=0.8\textwidth,keepaspectratio=true,angle=0}
}
\end{figure*}
\vskip 4mm

\subsection{The Fits}

Quasar continua are generally well-fit by a power law.  However,
superimposed on this smooth continuum is a ``pseudocontinuum'' of
broad \feii\ multiplet emission that effectively litters the entire
optical/UV region of the spectrum.  Although much progress has been
made deriving theoretical \feii\ spectra
\citep[e.g.,][]{verneretal2004}, the prospect of fitting $>800$
transitions individually is a daunting task.  Rather, it is common
practice to derive an \feii\ template from a high S/N observation of
an active galaxy whose broad emission lines are intrinsically narrow
($\lesssim 1000$~\kms), typically \zw.  Here we use the optical
template of \citet{borosongreen1992} and the UV template presented in
\citet{salvianderetal2007} that was derived from that of
\citet{vestergaardwilkes2001}, but has a theoretical template pasted
in beneath the \mgii\ line from \citet{sigutpradhan2003}.  Vestergaard
\& Wilkes also provide a separate \feiii\ template at $\approx
1900$~\AA\ that we utilize in attempting to fit the \ciii~$\lambda
1909$~\AA\ line.  In addition, we follow \citet{dietrichetal2002} and
model the ``little blue bump'' \citep{grandi1982} as a combination of
higher-order Balmer emission lines \citep{storeyhummer1995} and
bound-free Balmer continuum emission with a temperature $T_e =
15000$~K, electron density $n_e = 10^8 –- 10^{10}$~cm$^{-3}$, and
optical depth $0.1 \leq \tau_{\nu} \leq 2$ \citep[following
][]{grandi1982}.

In practice, we fit the power-law continuum, \feii\ emission, and
Balmer continuum simultaneously, using emission-line--free windows
(see Figure 1).  The fits include a normalization and slope for the
(single) power-law component, a width, shift, and amplitude for the Fe
template, and a density and optical depth for the Balmer continuum.
In the UV, we find the best results fixing the \feii\ broadening to
that of \hbeta, while in the optical we find negligible difference in
allowing the broadening to be free.  We subtract the continuum model
and create a line-only spectrum.

We fit the emission line spectra with multi-component Gaussians.
These components are used only to create a high-fidelity noise-free
match to the emission lines; we do not ascribe physical meaning to
them.  We first fit the \hbeta+\oiii~$\lambda \lambda 4959, 5007$
region.  The narrow \oiii\ lines are fit with up to three Gaussians
each, with the relative wavelengths tied to laboratory values and the
line intensities constrained to have a flux ratio of 1:3 \citep[Table
2; see details in][]{greeneho2005o3,greeneetal2009}.  The width of the
narrow \hbeta\ line (fit with a single Gaussian) is tied to that of
\oiii\ whenever possible (Table 3)\footnote{SBS0909+523 is the one exception.  
The narrow line component
is very strong in this object, even in the UV lines (see Figs. 1 \&
2) and is generally narrower in other transitions than in the \oiii\
line. Interestingly, the broad lines appear to be redshifted
compared to the narrow lines in this object.  It would be worth
attempting integral-field spectroscopy to investigate any spatial
offsets corresponding to the observed velocity offset.}.
Finally, broad \hbeta\
is fit with up to four (typically two) broad components.  We also find
the need to impose a lower limit of 1000~\kms\ on the broad components
so that they do not erroneously fit a narrow-line component.  We note
that in these bright quasars the flux contribution from the narrow
lines is typically small.  

\begin{figure*}
\vbox{ 
\hskip 0.6in
\psfig{file=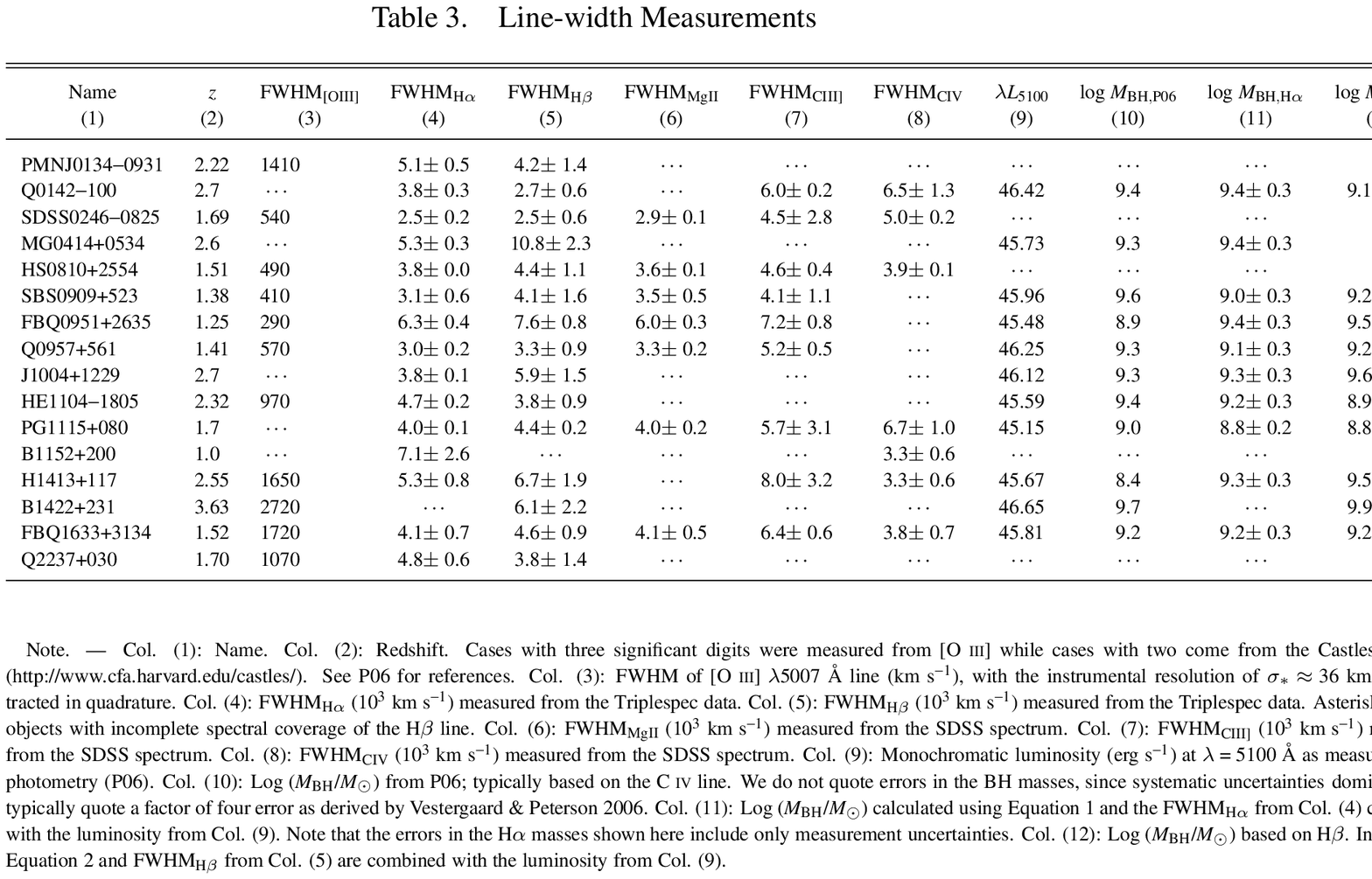,width=0.8\textwidth,keepaspectratio=true,angle=0}
}
\end{figure*}

\hskip -0.1in
\psfig{file=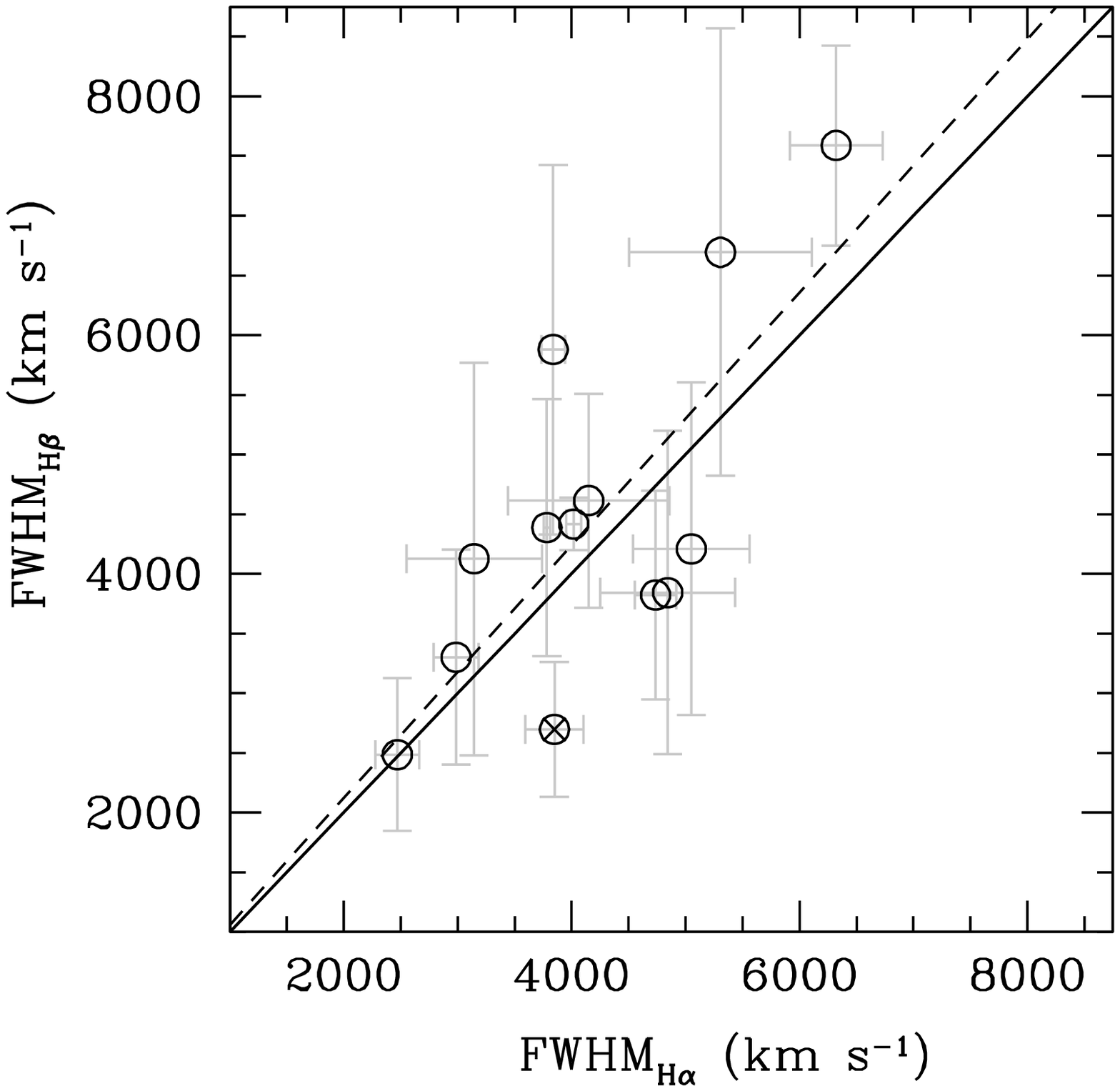,width=0.45\textwidth,keepaspectratio=true,angle=0}
\vskip -0mm
\figcaption[]{
Comparison between the line widths derived from \halpha\ and \hbeta\
in the Triplespec data.  \hbeta\ lines with incomplete spectral
coverage are noted with a cross.  The one-to-one line ({\it solid})
is shown for reference only.  As expected
\citep[e.g.,][]{greeneho2005cal} \hbeta\ is generally broader than
\halpha.  Excluding the compromised \hbeta\ fits, we find
$\langle$FWHM$_{\rm \halpha}$/FWHM$_{\rm \hbeta} \rangle = 0.9 \pm
0.2$ ({\it dashed line}), which is consistent with the $7\%$
difference in line width found by \citet{greeneho2005cal}, albeit
with considerable scatter.  We note that MG0414+0534 has a very
large and very unreliable \hbeta\ line that is off the scale of this
plot.
\label{cfhahb}}
\vskip 5mm
\noindent
Nevertheless, the treatment of the narrow
lines is a source of uncertainty in the FWHM measurements,
particularly in the UV.  As described below, we therefore perform an
additional fit with no narrow component and fold the 
difference into our total error budget.

Our approach is similar in the case of the \hn\ complex \citep[see
also][]{hoetal1997,greeneho2004}.  Here we fix the narrow \halpha\ and
\nii\ to the \oiii\ line width, fix the relative wavelengths to
laboratory values, and fix the relative strengths of the \nii~$\lambda
\lambda 6548, 6584$~\AA\ to 1:3.  In cases where there is no \oiii\
line, we fix the narrow-line width to 500~\kms.  Generally in these
cases the narrow components are too weak to be fit independently.
Uncertainties in this procedure are estimated using an alternate fit
with no narrow-line components.  We fit the broad \halpha\ with as
many as four Gaussians, but do not tie them to the \hbeta\ profile in any
way.  All of the Triplespec spectra and fits are shown in Figure 1 and the
Appendix.

Our fits to the UV lines proceed in a similar fashion, with each broad
line modeled as the sum of up to four Gaussians.  We still choose to
tie the narrow-line components to the width of \oiii\ as above,
although we note that the proper treatment of the narrow component of
\civ\ remains a matter of debate in the literature
\citep{baskinlaor2005,vestergaardpeterson2006,
  kellybechtold2007,shenetal2008b}.  Again, in cases without available
\oiii\ fits we have simply chosen a representative width of 500~\kms,
and again we perform a second, narrow-line--free, fit.  Finally, we
include a linear continuum component to remove residual continuum
errors (Fig. 2).  We note that the quality of the \mgii\ fit
does depend on the \feii\ subtraction, and our model for the \feii\
continuum is poorly constrained directly beneath the \mgii\ line.

\hskip -0.1in
\psfig{file=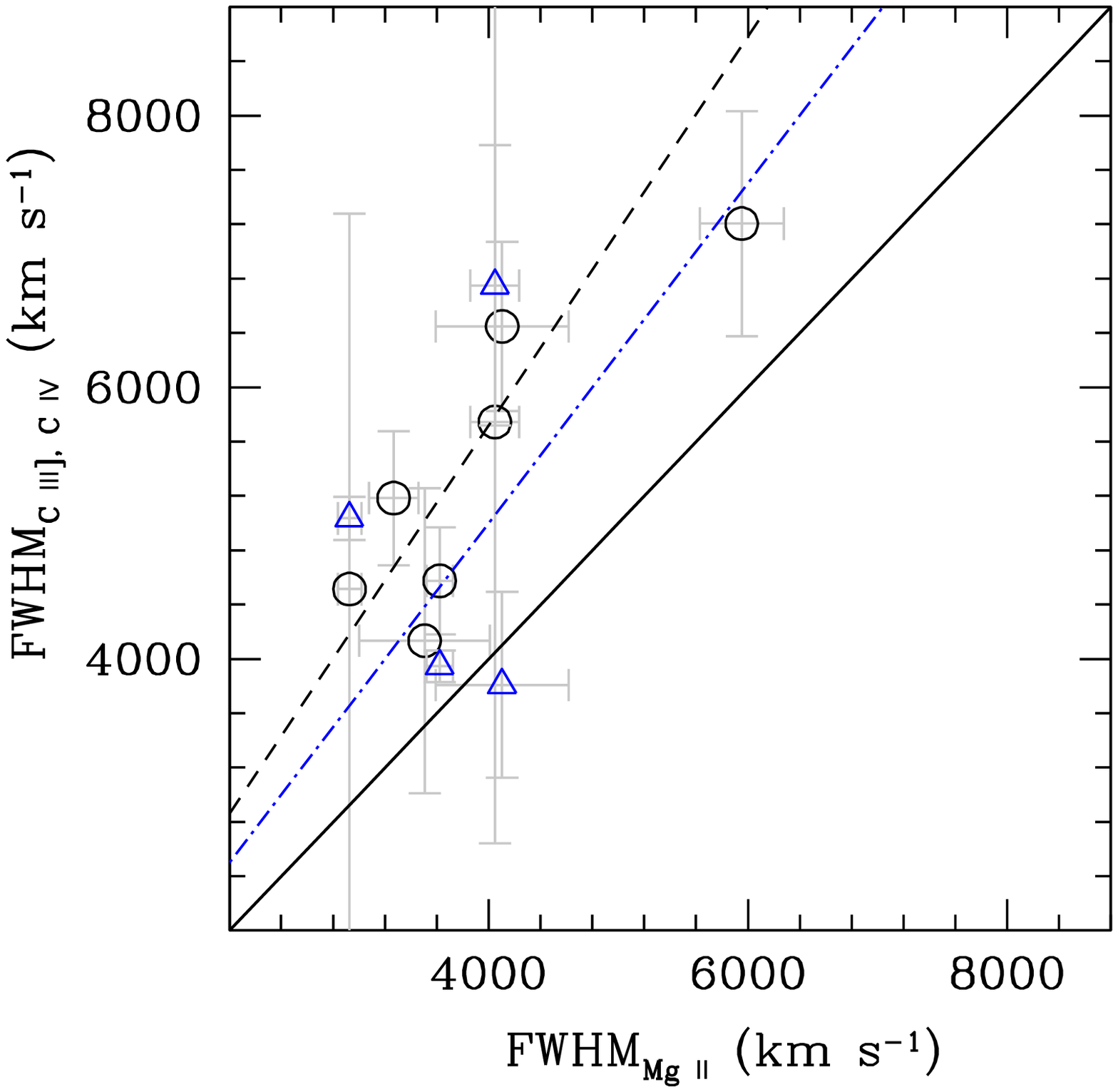,width=0.45\textwidth,keepaspectratio=true,angle=0}
\vskip -0mm
\figcaption[]{
Comparison between the line widths derived from Mg {\tiny II} and C
{\tiny III}] ({\it black open circles}) or C {\tiny IV} ({\it blue
  open triangles}) from the SDSS data.  The one-to-one line ({\it
  solid}) is shown for reference only.  We find
$\langle$FWHM$_{\rm{Mg {\tiny II}}}$/FWHM$_{\rm{C {\tiny III}]}}
\rangle = 0.7 \pm 0.1$ ({\it black dashed line}) and 
$\langle$FWHM$_{\rm{Mg {\tiny II}}}$/FWHM$_{\rm C {\tiny IV}} \rangle = 0.8 \pm 0.2$ 
({\it blue dot-dashed line}).
\label{cfmgc}}
\vskip 5mm
\noindent\
However, numerous authors have now shown that the uncertainty in FWHM
incurred by differing \feii\ models is small \citep[$<0.02$
dex;][]{salvianderetal2007,fineetal2008}.

In addition to the standard \mgii\ and \civ\ lines, we have attempted
to model the \ciii$~\lambda 1909$ transition, since it seems worth
investigating every possible transition in the UV [and at least in one
object, reverberation mapping with this line yielded a mass that is
consistent with other transitions; \citet{onkenpeterson2002}].  There
are good reasons to avoid \ciii, including blending with
\ion{S}{3}]$~\lambda 1892$, \ion{Al}{3}$~\lambda 1857$, and \feiii\
multiplets \citep[e.g.,][]{dietrichetal2002}.  To minimize these
degeneracies, each of the three lines (\ciii, \ion{S}{3}], and
\ion{Al}{3}) is modeled as a single Gaussian with the same width.  The
relative centroids are fixed to laboratory values and the strengths
are left free.  We include a narrow component only for \ciii, and it
is fixed to the \oiii\ width as above.  Although we also attempted to
include the \feiii\ multiplets \citep{vestergaardwilkes2001} in the
fit, keeping the width constrained to that of the \feii\ multiplets,
the results were not well-constrained.  It can be seen in Table 3
that, unexpectedly, in three out of six cases FWHM$_{\rm C {\tiny
    III}]}$ is larger than FWHM$_{\rm C {\tiny IV}}$.  However, these
are systems with narrow or broad absorption-line systems, which make
the \civ\ fits particularly uncertain.

After fitting, our velocity measure of choice is a non-parametric
full-width at half maximum (FWHM) derived from our multi-Gaussian
fits.  We are aware that many authors advocate the use of the line
dispersion rather than the FWHM
\citep[e.g.,][]{petersonetal2004,onkenetal2004,denneyetal2009line}.
We do not believe there to be a strong argument in favor of one or the
other line-width measurement at the current time \citep[although
see][]{collinetal2006}, but we do know that our technique is robust in
the presence of noisy spectra, even when fitting thousands of spectra.
We direct the interested reader to the Appendix of
\citet{greeneho2007c} for a detailed explication of our reasoning,
but see also \citet{denneyetal2009line}.

\hskip -0.1in
\psfig{file=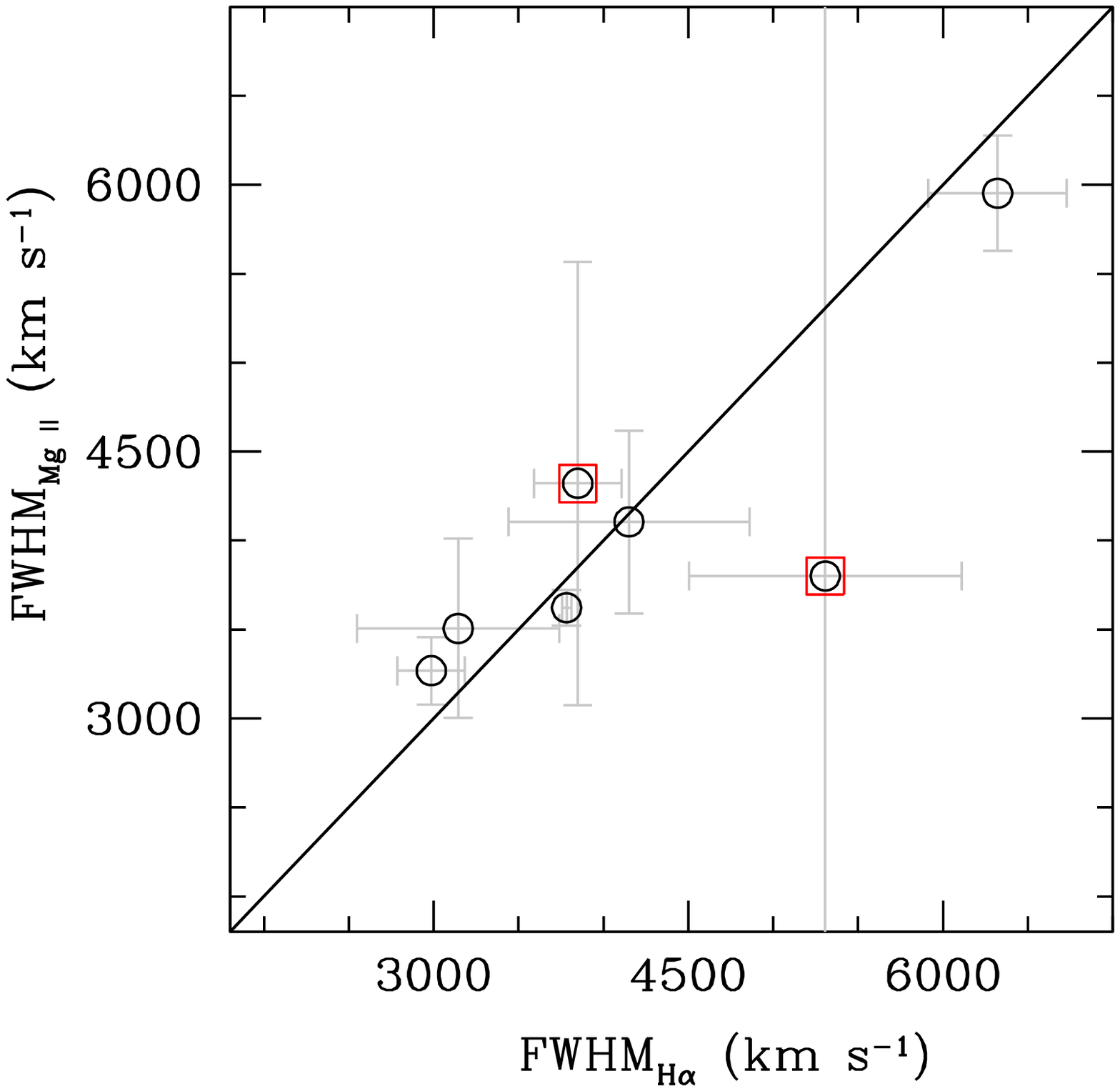,width=0.45\textwidth,keepaspectratio=true,angle=0}
\vskip -0mm
\figcaption[]{
Comparison between the line widths derived from \halpha\ and Mg
{\tiny III} or a scaled C {\tiny III}] using a ratio of 0.7 (Fig. 4)
when necessary ({\it red open squares}).  As above, we show the
one-to-one relation as a solid line.  Within the scatter, the two
measures agree, with $\langle$FWHM$_{\rm H\alpha}$/FWHM$_{\rm Mg
  {\tiny II}} \rangle = 1.0 \pm 0.2$.  
\label{cfhauv}}
\vskip 5mm

\subsection{Linewidth Comparisons}

In this subsection we present figures comparing
various estimates of line width derived from different elemental
transitions (Figs. 3--6).  It is clear from these figures that
measurement uncertainties alone lead to a significant amount of
scatter and that the total dynamic range in linewidth for any
transition is only a factor of $\lesssim$ three
\citep[e.g.,][]{fineetal2008,shenetal2008b}.

For each comparison figure, we have calculated the mean 
and standard deviation in the ratio of the two lines (not the error in
the mean).  These numbers are quoted in the figure captions.  We have
also calculated non-parametric correlation coefficients for each line
pair.  However, presumably because the samples are so small, none of
the correlations are formally significant.

We cannot come to any strong conclusions based on these comparisons,
due to both the small sample size and the large measurement errors.
We confirm that \hbeta\ is typically broader than \halpha\
\citep[Fig. 3; e.g.,][, and references therein]{greeneho2005cal}.  It
has long been known that high-ionization lines such as \civ\ are
typically broadened and blue-shifted compared to, e.g., \mgii\ and the
Balmer lines \citep[e.g.,][]{gaskell1982,osterbrockshuder1982,
  baldwinetal1996}.  While our observations are consistent with that
trend (Fig. 4), we cannot say much more than that.  Over this limited
range, we do not see strong evidence of a correlation between the
widths of the Balmer lines and \civ, but to a large degree the scatter
is driven by the difficulty in measuring a reliable width for the
broad-absorption system in H1413+117.  Furthermore, the (few) \ciii\
measurements we have seem to be correlated with the \mgii\ line widths
(Fig. 4).  We believe \ciii\ merits investigation with a much larger
sample.  Finally, we tend to derive somewhat broader line widths than
P06 from the UV lines, presumably because we have subtracted the
\feii\ emission, which P06 could not do (Fig. 6).  In a couple of cases the
difference in narrow-line treatment also plays a role.  We
plan to revisit the UV-based mass estimates for the entire sample
using modern spectroscopy (e.g., from SDSS) in future work.

\hskip -0.1in
\psfig{file=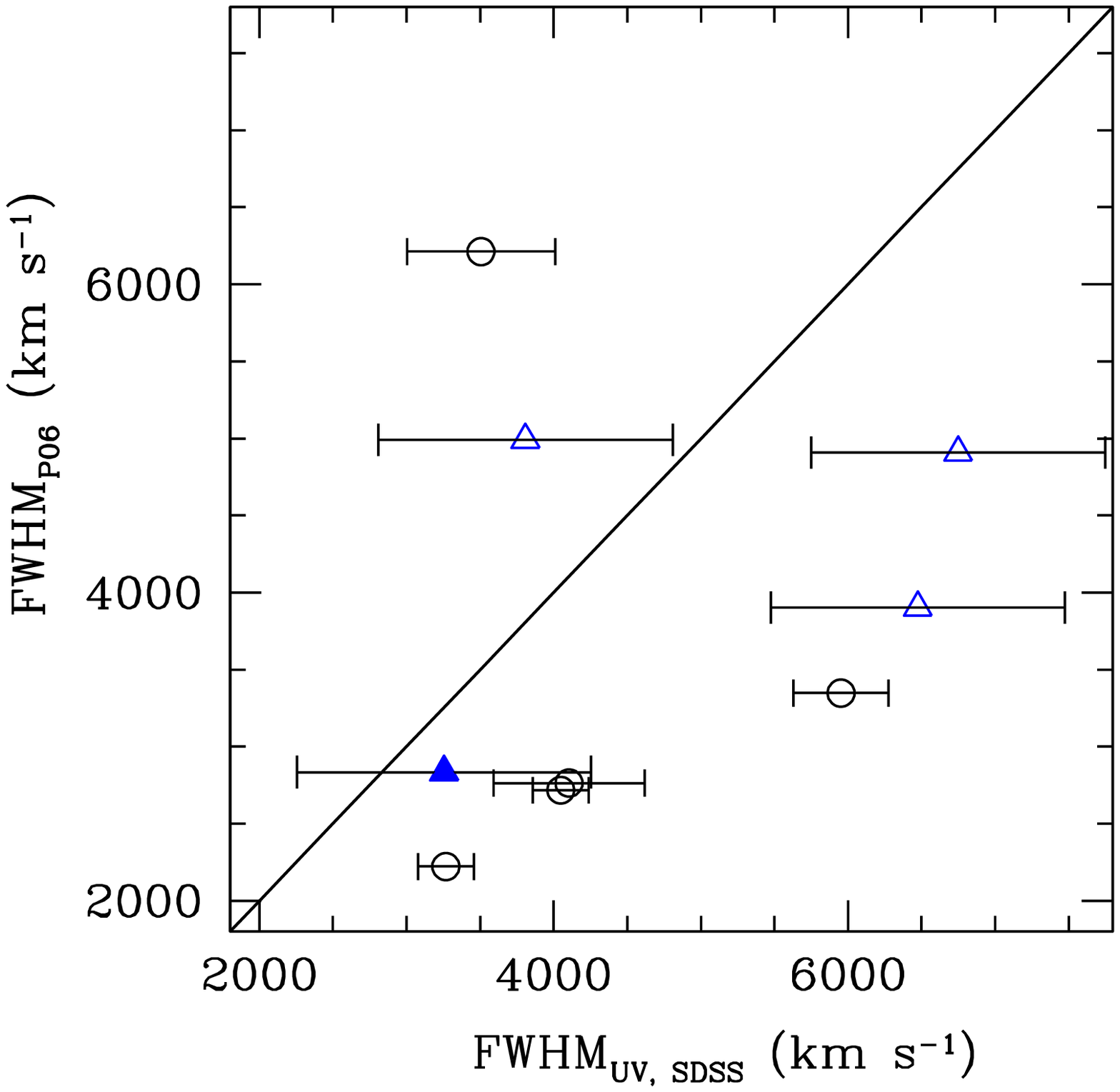,width=0.45\textwidth,keepaspectratio=true,angle=0}
\vskip -0mm
\figcaption[]{
Comparison between our measured UV line widths from Mg {\tiny II}
({\it black open circles}) or C {\tiny IV} ({\it blue open triangles})
and those published in P06.  Again, the scatter is quite large
$\langle$FWHM$_{\rm Mg {\tiny II}}$/FWHM$_{\rm P06} \rangle = 1.5
\pm 0.5$, and again the unity relation is shown as a solid line.
Our tendency to derive broader line widths than P06 derives from our
continuum subtraction and treatment of the narrow lines.  H1413+117
is highlighted as a filled square because the broad absorption
feature in the C {\tiny IV} line makes fitting difficult.
\label{cfhap06}}
\vskip 5mm

\subsection{Mass Estimates}

The primary goal of our study is to remove any potential systematic
bias in the \civ-based masses presented in P06 by calculating
Balmer-based virial masses.  In calculating \halpha-based BH masses,
we start with the radius-luminosity relation of
\citet{bentzetal2009a}.  Since this relation was calibrated using
\hbeta, we then convert \fwhb\ to \fwha\ using the relation derived in
\citet{greeneho2005cal}, which both increases the scaling with \fwha\
to the 2.06 power and slightly changes the prefactor.  Finally, for
consistency with P06, we assume a scaling factor that is $1.8$ times
higher than the assumption of isotropic random motions,
\citep[e.g.,][]{onkenetal2004,greeneho2006msig} yielding:

\hskip -0.6in
\begin{equation}
\mbh = (9.7 \pm 0.5) \times 10^6 
\left(\frac{\lf}{10^{44}~{\rm erg~s^{-1}}} \right)^{0.519 \pm 0.07} 
\left(\frac{\fwha}{10^3~{\rm km s^{-1}}} \right)^{2.06 \pm 0.06}
\end{equation}

The masses calculated using this formalism are displayed in 
Table 3 and Figure 7{\it a}. Note that in the case of B1422+231, 
\halpha\ is outside of our observing window.  In this case, we 
use the \hbeta\ line, and the following (very similar) formalism:

\hskip -0.6in
\begin{equation}
\mbh = (9.1 \pm 0.5) \times 10^6 
\left(\frac{\lf}{10^{44}~{\rm erg~s^{-1}}} \right)^{0.519 \pm 0.07} 
\left(\frac{\fwhb}{10^3~{\rm km s^{-1}}} \right)^2.
\end{equation}
\noindent
The \hbeta\ masses are presented in Figure 7{\it b}.

The luminosities used here are derived from the \hst\ photometry of
the CASTLES sample, as described in detail in P06.  Briefly, the
quasar and host galaxy images are optimized along with the lens model
using LENSFIT \citep[a variant of GALFIT;][]{pengetal2002}.  The
quasars are modeled as point sources, the galaxies as \sers\
(\citeyear{sersic1968}) functions, and the lensing masses as singular
isothermal ellipsoids.  The demagnified and deblended $V, I, H$ quasar
magnitudes are modeled as a power-law, with the slopes presented in
Table 2, and the luminosities come from this fit.  We note that our
spectroscopically derived slopes agree nicely with those from
broad-band photometry.  We find $\langle \alpha_{\rm opt}/\alpha_{\rm
  P06}\rangle = 1 \pm 0.5$, where $f_{\lambda} \propto
\lambda^{\alpha}$.  The \hst\ $H$-band images correspond closely to
restframe $V$-band at $z \approx 2$, so the luminosities are very
insensitive to uncertainties in the powerlaw slope.

We find a mean ratio of $\langle {\rm log} (M_{\rm H\alpha}/M_{\rm P06})
\rangle = -0.02 \pm 0.5$ between the two mass estimates.  Although the 
scatter is large, we find no evidence for a systematic offset
in the \civ-based masses.  Therefore, to the extent that virial mass
estimates have merit in this mass, luminosity, and redshift regime (as
yet untested directly), our new spectroscopic observations confirm the
results presented in P06.

\subsection{Uncertainties}

Uncertainties in line widths are difficult to estimate.  In our case,
the signal-to-noise (S/N) ratios of the spectra are not very high (see
Table 1), which contributes substantially to the error budget.  We use
Monte Carlo simulations to estimate the magnitude of the uncertainties
due to finite S/N.  For each observed line, we generate 1000 mock
spectra with the same shape as our best-fit model and the same noise
as our observed spectrum.  In order to model residual uncertainties in
continuum subtraction, we include a broadened \feii\ spectrum with an
amplitude that scatters around $10\%$ of the peak flux.  Using the
same procedures as in the real data, we then measure the best-fit
parameters for each of these simulations.  The uncertainty is
determined from the distribution of simulated line dispersion to be
half of the width encompassing 68\% of the simulated galaxy
measurements, and ranges from $\sim 10-40\%$ of the measured value.
Now, in addition to noise, the treatment of narrow emission lines can
be a significant cause of systematic uncertainty.  Thus we perform a
second fit with the narrow component of each line turned off, and the
difference between the two represents a second estimate of the
uncertainty.  We adopt the larger of these as our final error in line
width.

At the same time, quasar variability adds additional uncertainty since
the rest-frame optical and UV spectra analyzed here were not obtained
contemporaneously.  \citet{wilhiteetal2007} and
\citet{denneyetal2009line} both find that variability in line width
tends to be small ($\sim 30\%$) and thus only contributes $\sim$ 0.1
dex scatter to BH mass estimates.  Another feature of quasar spectra
that complicates emission-line--width measurements is the presence of
blue-shifted absorption features with velocity widths of tens (narrow)
to tens of thousands (broad) of \kms.  Both H1413+117 and (to a lesser
extent) PG1115+080 display broad absorption features, while Q0957+561
and HS0810+2554 have significant narrow absorption.  In these cases we
simply mask the regions from the fit, but note that particularly in
the case of H1413+117, our ability to measure a reliable \civ\ width
is seriously compromised, since even the line center is not
well-defined.

In addition to the line widths presented here, the BH masses depend on
the luminosity of the quasar.  We briefly review the arguments that
the lens modeling does not add significant uncertainties to our
results and direct the interested reader to Appendix B of P06 for more
details.  The primary source of uncertainty in the models is that
gravitational time delay and lensing substructures can lead to
anomalous quasar magnification ratios, which translate into an
uncertainty in the quasar luminosity.  However, this error should be
no larger than 0.1-0.2 mag, and thus does not contribute significantly
to the BH mass uncertainty.  Actual measurement uncertainties from
model fitting are tiny for the quasar.

In Table 3 we present the formal uncertainties in BH mass arising from
errors in the FWHM measurements and the formal uncertainty in the
slope of the radius-luminosity relation.  We wish to emphasize that
the true errors in BH mass are probably dominated by systematic
uncertainties arising from our ignorance of the structure and
kinematics of the broad-line region that translate into errors in
inferring both its true extent and velocity field based on the
observations \citep{krolik2001,collinetal2006, greeneho2006msig}.  At
present, we really have no concrete confirmation that (a) the same
radius-luminosity relation applies to quasars at these luminosities
\citep[although see][]{kaspietal2007} nor that (b) it is meaningful to
assume that the BLR gas is in virial equilibrium and not, for
instance, dominated by a massive outflow
\citep[e.g.,][]{baldwinetal1996,richardsetal2002a,leighlymoore2004,
  fineetal2008}.

\begin{figure*}
\vbox{ 
\vskip -0.1truein
\hskip -0.2in
\psfig{file=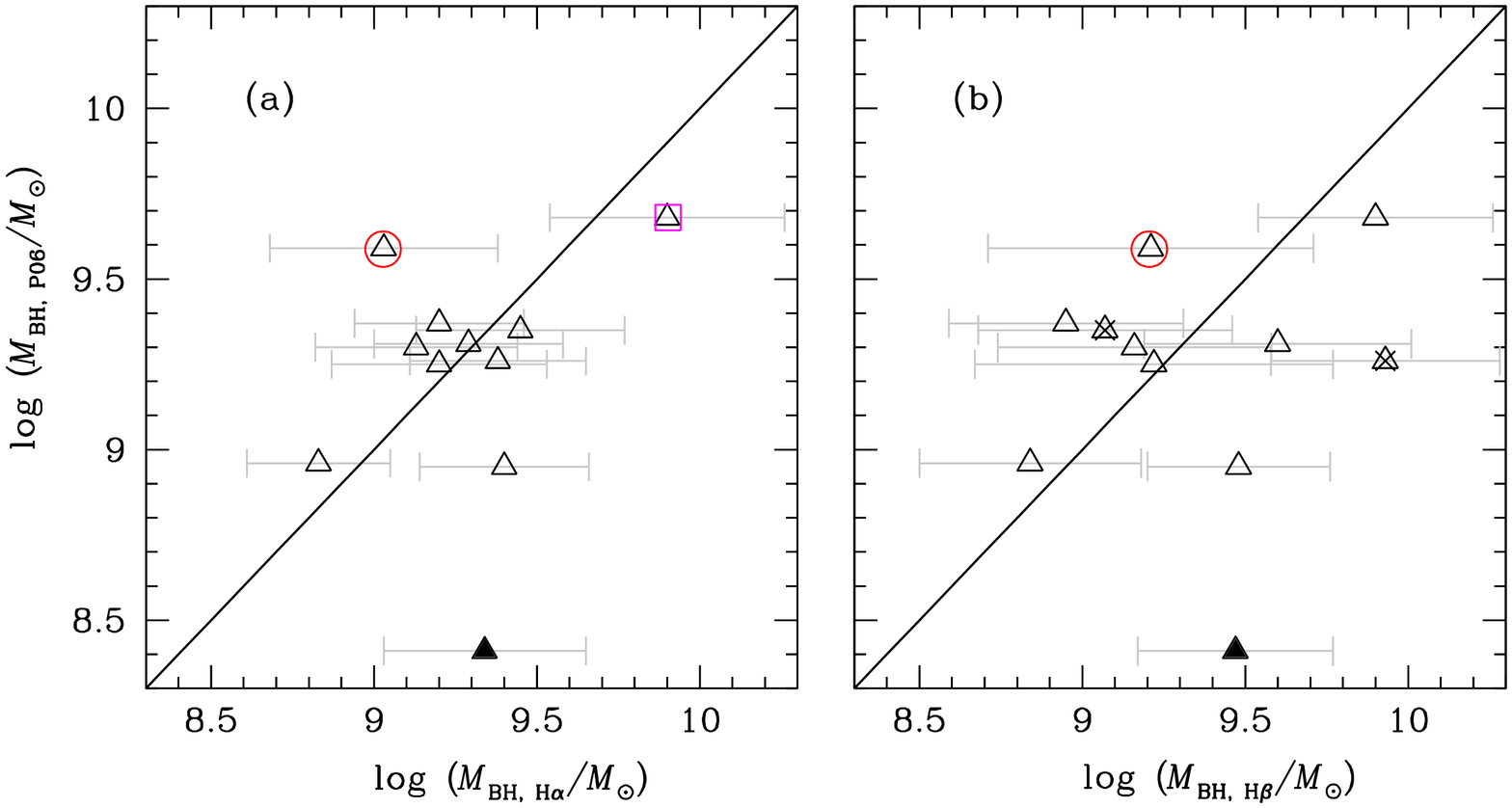,width=0.55\textwidth,keepaspectratio=true,angle=-90}
}
\vskip -0mm
\figcaption[]{
({\bf a}): We compare BH mass estimates based on the \halpha\ 
transition with those from P06.  The P06 masses are based on C {\tiny IV} 
with one exception ({\it red open circle}; Mg {\tiny II}) while our masses 
are based on \halpha\ with one exception 
({\it magenta open square}; \hbeta).  The unity relation ({\it solid line}) 
is shown to guide the eye.  While the two mass estimates are not strongly 
correlated, we do not find any evidence for a systematic offset between 
the two (the median $M_{\rm H\alpha}/M_{\rm P06} = 1.0 \pm 0.4$).
H1413+117, due to its broad-absorption system, is highlighted 
as a filled symbol.
({\bf b}):  Same as (a), except using \hbeta\ as the virial indicator.  
In this case objects with only partial observations of the \hbeta\ 
line are highlighted with crosses.  As above the P06 mass that is 
based on Mg {\tiny II} is identified with a red circle. We find 
a median $M_{\rm H\beta}/M_{\rm P06} = 1.0 \pm 0.5$.  The filled 
triangle is H1413+117.
\label{cfmass}}
\end{figure*}
\vskip 5mm

\section{Demographics of Lensed Quasars}

\begin{figure*}
\vbox{ 
\vskip -3mm
\hskip -0.2in
\psfig{file=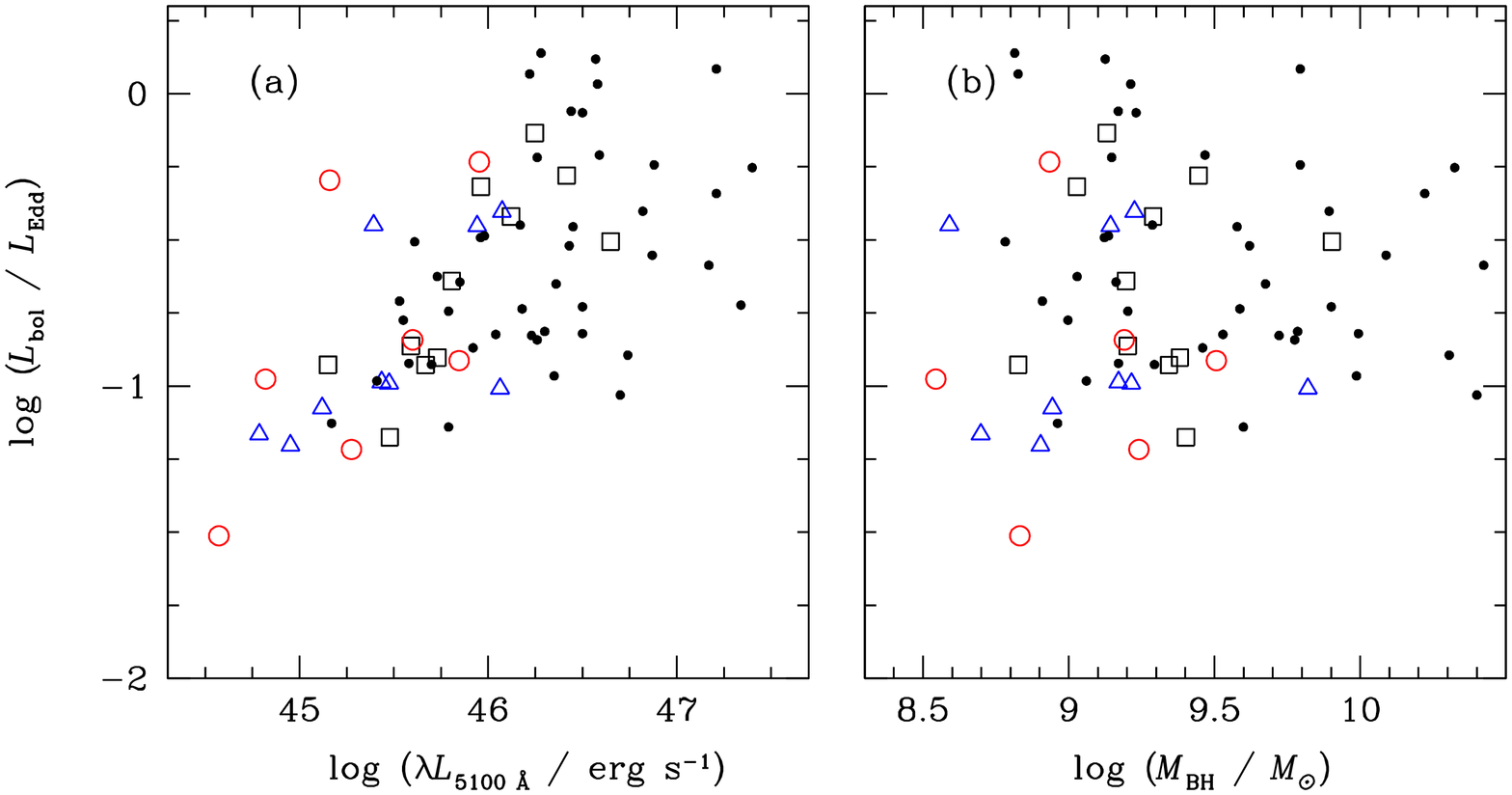,width=0.55\textwidth,keepaspectratio=true,angle=-90}
}
\vskip -0mm
\figcaption[]{
{\bf (a)}: We reproduce Figure 2 from \citet[][{\it black dots;
  based on \hbeta}]{netzeretal2007} and include the targets from
this study ({\it large black open squares}), as well as the entire
P06 sample, differentiating between Mg {\tiny II}-based ({\it small red
  circles}) and C {\tiny IV}-based ({\it small blue triangles}) masses.
Here we plot monochromatic 5100 \AA\ luminosity versus the Eddington
ratio.  We follow Netzer et al. and assume that the bolometric
luminosity is seven times \lf.  Both our mass estimates and those
from Netzer et al. are based on Balmer emission lines (\hbeta\ in
their case).
{\bf (b)}: \mbh\ versus \lledd; symbols as in {\bf (a)}.  
We see that the distributions 
in each diagram are similar between the two samples, although the 
lensed systems extend to objects that are factors of two to three 
times fainter.  
\label{fits}}
\end{figure*}

To summarize, we have obtained \halpha-based BH masses for a large
fraction of the lensed quasars with measured host galaxy luminosities
from P06.  While the scatter between \halpha-based and \civ- or
\mgii-based BH masses is quite large ($\sim 0.5$ dex), we do not see
evidence for a systematic bias in mass.  Thus, we do not alter the result of
P06 that \mbulge\ relations appear to evolve with cosmic time.

It is instructive to now compare the properties of the lensed sample
with the general quasar population at $1 \lesssim z \lesssim 4$.  A
large number of studies have looked at the distributions of BH mass
and Eddington ratio of luminous quasars
\citep[e.g.,][]{mcluredunlop2004,kollmeieretal2006,netzeretal2007,
  fineetal2008,shenetal2008b,gavignaudetal2008} including some work on
narrow-line quasars \citep{greeneetal2009,liuetal2009} as well as mass
functions for quasars
\citep[e.g.,][]{greeneho2007b,vestergaardetal2008,kellyetal2009}.  In
most cases the BH masses are based on \civ\ and \mgii\ from observed
optical spectra.  Generally, all works find the same surprisingly
narrow range in measured line width, and correspondingly narrow
distributions in \mbh\ and \lledd\ at all redshifts (although, as
Gavignaud et al. point out, the derived distribution is quite
sensitive to the assumed slope in the radius-luminosity relation).

For our purpose it is easiest to compare with the results of Netzer et
al., as their study is also based on \hbeta\ observations.  For
convenience we adopt their assumption that \lf\ is $\sim$ one seventh
of the bolometric luminosity and we take the Eddington luminosity to
be $1.26\times10^{38}$~(\mbh/\msun) \lum.  Despite completely
different selection criteria, the distributions in luminosity,
and correspondingly inferred BH mass and Eddington ratio, are quite
similar across the two samples (Fig. 8).  For the Triplespec sample,
the median $\langle$\mbh$\rangle \approx 2 \times 10^9$~\msun and the
median \lf$\approx 6 \times 10^{45}$~erg~s$^{-1}$ yield a typical
Eddington fraction of $\sim 30\%$.  If we take the entire P06 sample,
using \halpha-based masses when available, we find a median mass of
$\sim 10^9$~\msun, a median \lf\ of $2 \times 10^{45}$~erg~s$^{-1}$,
and a median Eddington ratio of $10\%$.

Unlike other studies, Netzer et al. actually argue that their sample
displays a broad distribution in Eddington ratio when compared to
theoretical expectations that luminous quasars should be most readily
observed at or near their Eddington luminosities
\citep[e.g.,][]{marconietal2004,merloni2004,hopkinsetal2006}.  We find
the observed distributions unexpected for a somewhat different reason,
namely the uniformly high BH masses.  Assuming that the Eddington
limit strictly applies and given the lower flux limit of the lensed
sample we could detect BHs with masses as low as $\sim 10^8$~\msun.
Furthermore, our intuition from the local BH mass function
\citep[e.g.,][]{yutremaine2002,marconietal2004} suggests that
$10^8$~\msun\ BHs ought to be far more numerous than $10^9$~\msun\
systems due to the exponential decline in the space density of the
most luminous galaxies.  Given that supercritical accretion may be
observed locally
\citep[e.g.,][]{poundsetal1995,mineshigeetal2000,desrochesetal2009},
we find the preponderance of $\sim 10^9$~\msun\ BHs radiating at $\sim
10\%$ of their Eddington luminosity to be noteworthy.

We can be slightly more quantitative.  Following, e.g.,
\citet{somerville2009}, we can transform the observed $z \sim 2$
galaxy mass function \citep[e.g.,][]{fontanaetal2006} into the
expected BH mass function.  For reference, at the median luminosity of
the Triplespec sample, the Eddington limit sets a bound of $\sim 4
\times 10^8$~\msun\ on observable BHs.  If the relation between \mbh\
and galaxy mass were identical to what it is today
\citep[e.g.,][]{gultekinetal2009}, then the BHs with mass $\approx 4
\times 10^8$~\msun\ ought to be $\sim 50$ times more common than the
median observed mass of \mbh$\approx 2 \times 10^9$~\msun.  Note that
we are conservatively adopting the median rather than the minimum
observed luminosity and assuming that the Eddington
luminosity applies strictly.  Taking the observed masses at face value
for the moment, and assuming that the selection of lensed quasars is
simply a random selection of optically luminous quasars at $1 < z <
4$, then we infer that either BHs radiating at 30\% of Eddington are
$\sim 50$ times more numerous than Eddington-limited objects or that
the zeropoint in the \mbulge\ relation has evolved by a factor of
$\sim 3$ to the present day.  This latter, of course, is the
suggestion made by P06.  

Obviously, such constraints are not particularly stringent at the
moment.  For one thing, the selection of lensed quasars is complicated
to model, involving radio selection in some cases \citep[which will
tend to bias samples toward more massive systems; e.g.,][]
{heckman1983,mandelbaumetal2009}.  Furthermore, as pointed out by
Somerville, joint constraints from galaxy and quasar luminosity
functions on their own cannot distinguish between zeropoint evolution
or increased scatter in BH-bulge relations at high redshift.  Finally,
we do not know that the quasar duty cycle is independent of mass.
Nevertheless, we cannot help but wonder whether the narrow range in
observed line width, and the unexpectedly high average BH mass, do not
instead indicate a problem in virial mass estimators at high
luminosity that may be resolved with better understanding of the
physics of broad-line quasars.

Ultimately, the question is whether or not the broad emission lines in
luminous quasars are dominated by virial motions.  Of course, there is
evidence for a non-virialized component in the \civ\ line
\citep[e.g.,][]{baldwinetal1996,richardsetal2002a}. On the other hand,
there is clear evidence for virialized motion in the broad-line
regions of at least a few nearby lower-luminosity active galaxies
\citep[e.g.,][]{petersonwandel1999,onkenpeterson2002,petersonetal2004}.
Also, the local virial masses seem to correlate with both \sigmastar\
\citep[e.g.,][]{shenetal2008a} and $L_{\rm bulge}$
\citep[e.g.,][]{kimetal2008b,bentzetal2009c}.  Unfortunately, similar
analysis is not yet available in the luminosity range of interest to
us.  At present, all we can say is that the \civ-based masses alone
are not causing a net bias in the virial masses relative to the Balmer
lines.  True evolution in BH-bulge scaling relations is by no means
certain; far more pernicious sources of uncertainty remain, including
potential biases in sample selection.  At the minimum, to mitigate 
these concerns, quasar samples with identical selection at multiple 
redshifts are needed.

In closing, we note that BH-bulge relations are not the only ones
purported to display unexpected evolution since a redshift $z \approx
2$.  Recent work has shown that, at a fixed mass, elliptical galaxies
were a factor of two to five smaller at redshifts of one and two
respectively than they are today \citep[e.g.,][]
{trujilloetal2006,vandokkumetal2008,franxetal2008,
  vanderweletal2008,damjanovetal2009}.  At first it seems
counterintuitive that the most massive elliptical galaxies, which seem
to have formed the bulk of their stars rapidly at an earlier epoch
\citep[e.g.,][]{thomasetal2005}, should grow less dense with time.
However, a large number of minor mergers can very efficiently build
the outskirts of elliptical galaxies
\citep[e.g.,][]{boylan-kolchinma2007,bezansonetal2009,naabetal2009}.
Interestingly, these gas-free minor mergers would need to grow the
total galaxy mass by factors of two to three in order to match local
observations \citep[e.g.,][]{bezansonetal2009}.  This is possible
under a scenario where the BH-to-TOTAL stellar mass relation is
steeper than linearity in low mass galaxies, which is observed in
nearby galaxies \citep{gultekinetal2009,greeneho2008}.  In gas-free
merging there is no corresponding BH growth via accretion, which could
facilitate a boost in the BH-bulge ratio at late times \citep[see also
][]{hopkinsetal2009}.  As argued by \citet{peng2007}, numerous minor
mergers will tend to drive the ratio of BH to galaxy mass towards a
value of unity slope (e.g., his Fig. 2b and 4a). Alternatively, as
suggested by both \citet{croton2006} and \citet{jahnkeetal2009}, it
may be that disk components are common in the high redshift galaxies
but are subsequently subsumed into the main elliptical galaxy body.
We mention this apparent coincidence in passing because it is
intriguing, although we note that a variety of outstanding
uncertainties are yet to be explored as far as the structural
evolution of elliptical galaxies is concerned.

\section{Summary}

We revisit the BH mass estimates for a sample of lensed quasars with
high-fidelity host-galaxy luminosities from \citet{pengetal2006b}.
While the Balmer-based masses presented here are arguably
more robust than the UV-based estimates, we find no evidence for a
systematic difference in BH masses based on the two methods.  If we
can take the Balmer-based virial masses at face value, then we confirm
the result of \citet{pengetal2006b} that BHs appear to be overly
massive relative to their hosts at high redshift.  Intriguingly, the
minor mergers that are invoked to puff up compact elliptical galaxies
at late times would be very effective at boosting the galaxy to BH
mass ratio as well.  However, as pointed out many times, the persistently
narrow range in observed line width for luminous quasars at high
redshift provides substantial cause for concern in the veracity of the
virial masses.  Furthermore, it is not clear that we have yet
assembled consistent comparison quasar samples across cosmic time with
which to compare the BH-bulge scaling relations.  Our work is a
necessary, but not sufficient, step in determining the true evolution
of BH-bulge relations with cosmic time.

\acknowledgements{The referee provided a very careful and insightful
reading that substantively improved the manuscript.  Both
J.~E.~G. and C.~Y.~P. are grateful for many stimulating
conversations with L.~C.~Ho and we thank A.~J.~Barth and C.~A.~Onken
for very useful comments.  We thank M. Skrutskie for taking
commissioning data for this project and for invaluable help with the
reduction software.  We thank H. Netzer for kindly providing his
data table for Figure 8.  Support for J.~E.~G. was partially
provided by NASA through Hubble Fellowship grant HF-01196 awarded by
the Space Telescope Science Institute, which is operated by the
Association of Universities for Research in Astronomy, Inc., for
NASA, under contract NAS 5-26555.}

\appendix{Triplespec Spectra}

In Figure A1({\it a-d}), we present the rest of the Triplespec spectra 
for completeness.  Note that the Triplespec spectra for SBS0909 and 
HS0810 are shown in Figure 1.

\begin{figure*}
\psfig{file=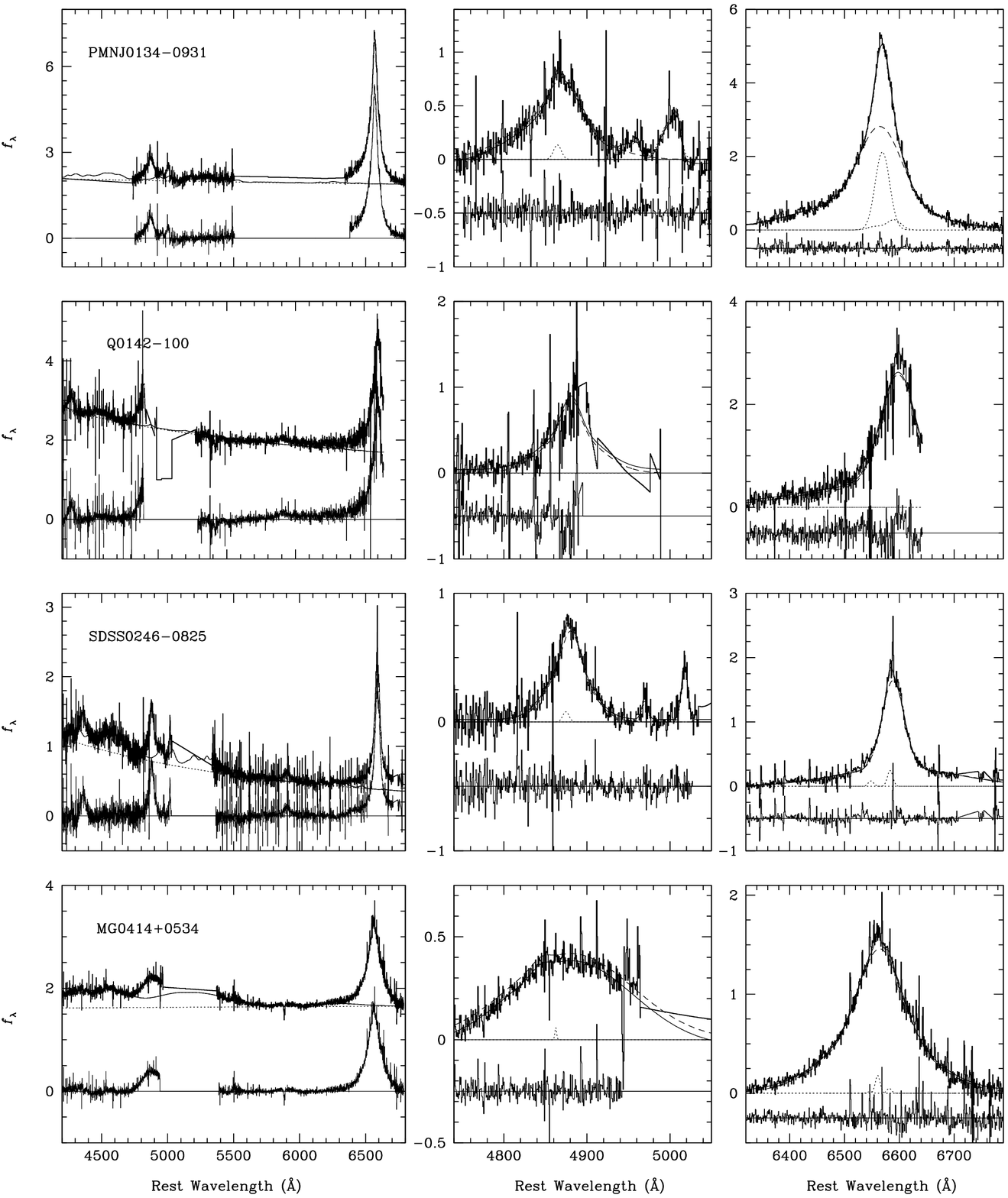,width=0.9\textwidth,keepaspectratio=true,angle=0}
\end{figure*}
\begin{figure*}
\psfig{file=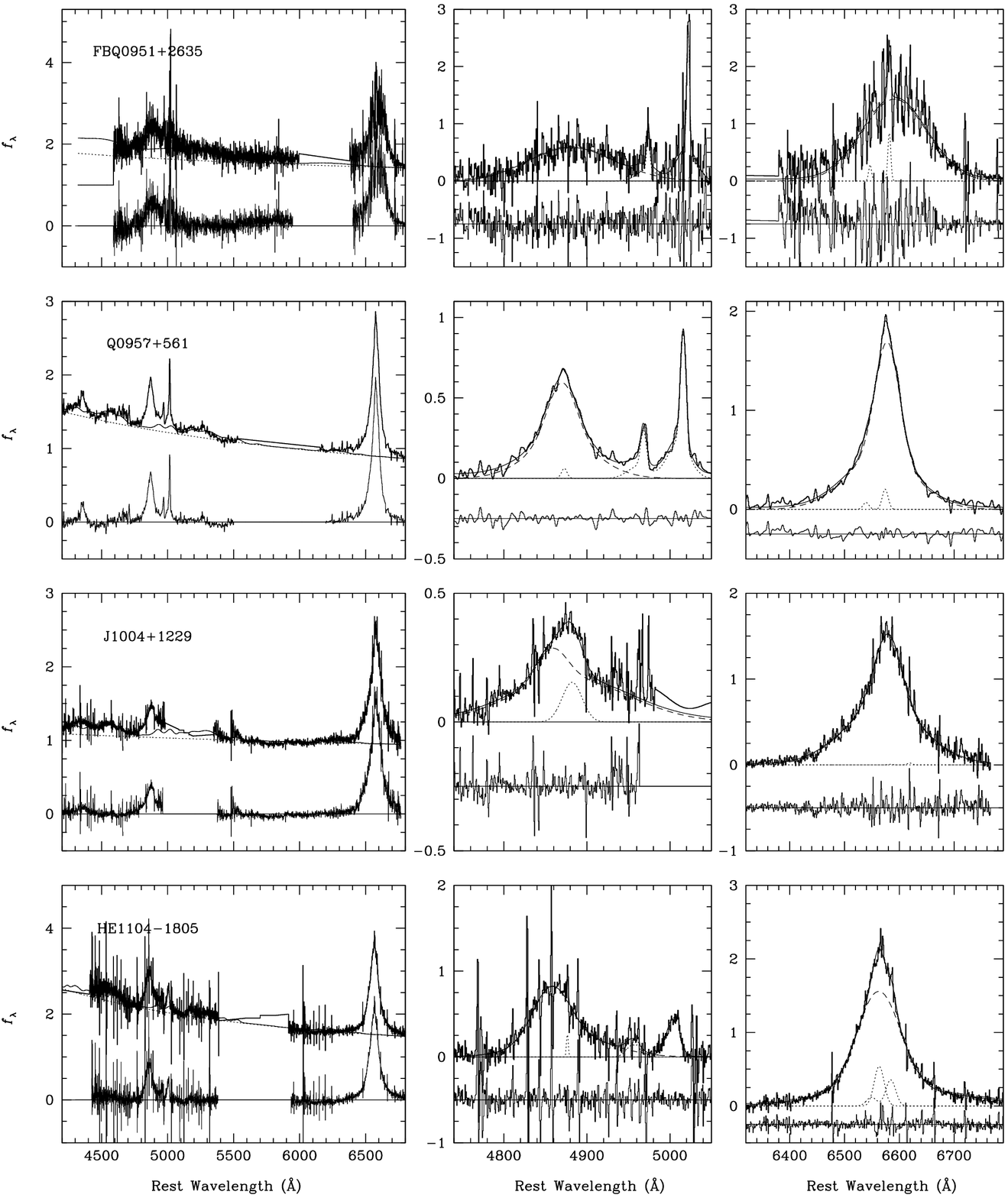,width=0.9\textwidth,keepaspectratio=true,angle=0}
\end{figure*}
\begin{figure*}
\psfig{file=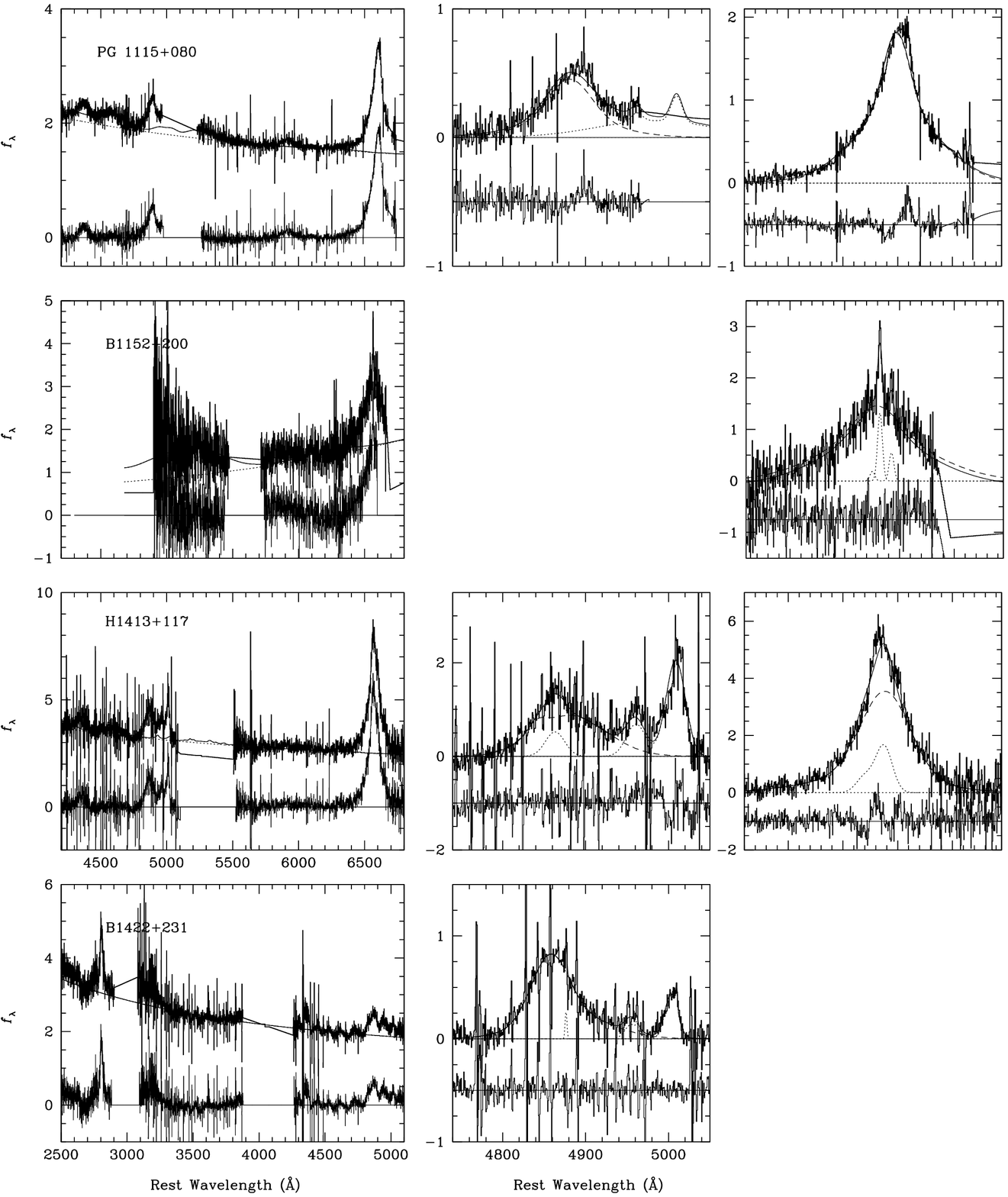,width=0.9\textwidth,keepaspectratio=true,angle=0}
\end{figure*}
\begin{figure*}
\vbox{ 
\vskip -25mm
\hskip 0.in
\psfig{file=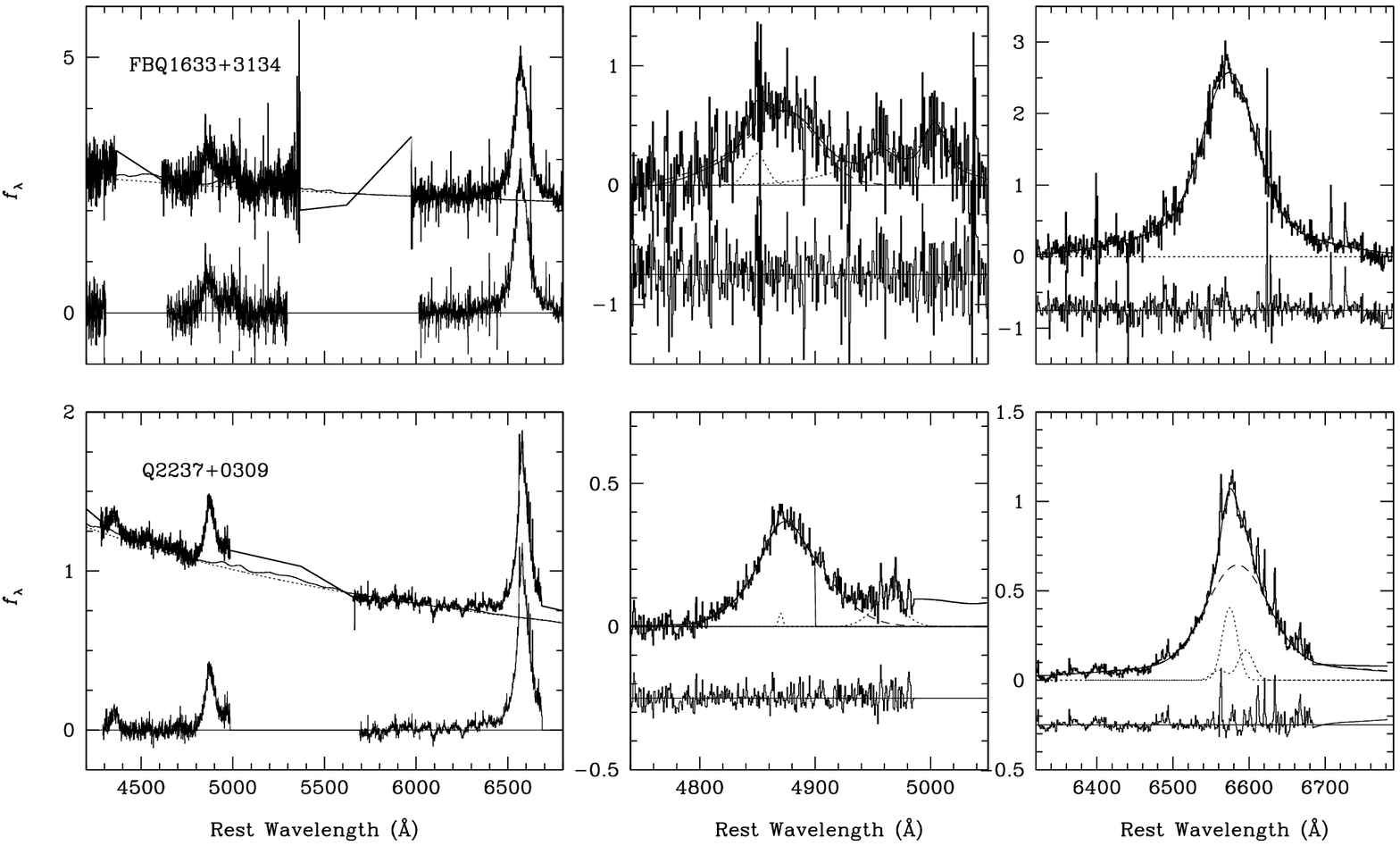,width=0.9\textwidth,keepaspectratio=true,angle=0}
}
\vskip -0mm
\figcaption[]{
Fits to the continuum ({\it left}), \hbeta\ ({\it middle}), 
and \halpha\ ({\it right}) lines from the Triplespec data that do not appear 
in Figure 1.  We show the 
data ({\it solid histogram}), the total model ({\it thin solid}), 
the broad- ({\it dashed}) and narrow-line ({\it dotted}) model 
components, and residuals below ({\it thin solid histogram}).  Data are 
plotted with an arbitrary scale in $f_{\lambda}$.  Spectral regions that are 
masked in the fit do not appear in the residuals (e.g., the red wing of 
\hbeta\ for Q0142-100).  Also note that in a couple of cases the [O {\tiny III}]
fit is based exclusively on the $\lambda 4959$ line; no redshifts are derived 
for these targets.
\label{appendix}}
\end{figure*}


\end{document}